\documentclass[12pt]{iopart}
\usepackage{amssymb}
\usepackage{bm}
\usepackage{graphicx}
\usepackage{epstopdf}
\expandafter\let\csname equation*\endcsname=\relax
\expandafter\let\csname endequation*\endcsname=\relax
\usepackage{amsmath}
\newcommand{\be}{\begin{equation}}
\newcommand{\en}{\end{equation}}
\newcommand{\prl}{Phys. Rev. Lett. }
\newcommand{\pre}{Phys. Rev. E }

\newcommand{\avg}[1]{\left< #1 \right>}

\newcommand{\tilh}{\tilde{H}}

\begin{document}
\title{Statistics of eigenvectors in the deformed Gaussian unitary ensemble of random matrices}
\author{K. Truong, A. Ossipov}
\address{School of Mathematical Sciences, University of Nottingham, Nottingham NG7 2RD, United Kingdom}

\begin{abstract}

We study eigenvectors in the deformed Gaussian unitary ensemble of random matrices $H=W\tilh W$, where $\tilh$ is a random matrix from Gaussian unitary ensemble and  $W$ is a deterministic  diagonal matrix with positive entries. Using the supersymmetry approach we calculate analytically the moments and the  distribution function of the eigenvectors components for a generic matrix $W$.  We show that  specific choices of $W$ can modify significantly the nature of the eigenvectors changing them from extended to critical to localized.  Our analytical results are supported by numerical simulations. 

\end{abstract}

\maketitle
\section{Introduction}
The Random Matrix Theory (RMT) was initially introduced in physics as a tool for  understanding energy levels of heavy atoms. The simplest and the most prominent ensemble of random matrices is the Gaussian Unitary Ensemble (GUE), in which elements $H_{ij}=H_{ji}^{\ast}$ of an $N\times N$ random Hermitian matrix are given by the independent Gaussian distributed complex random variables with the zero mean value $\avg{H_{ij}}=0$ and the variance  $\avg{|H_{ij}|^2}=1/N$. This ensemble captures many universal features of complex quantum systems and has numerous applications in other fields \cite{RMT_book}.

While the statistical properties of the eigenvalues of GUE matrices are highly non-trivial, the statistical distribution of their eigenvectors is very simple. The normalized eigenvectors are distributed uniformly  over a unit sphere in $\mathbb{C}^N$. In the limit $N\to \infty$, all the eigenvector components $\psi_n$ become independent and the distribution function of $y=N|\psi_n|^2$ is given by 
\be\label{gue_distr}
P(y)=e^{-y},
\en 
which in particular implies that  $\sum_{n=1}^N\avg{|\psi_n|^4}\propto 1/N$. The quantity $I_2=\sum_{n=1}^N\avg{|\psi_n|^4}$ is known as an inverse participation ratio and it measures the number of components contributing significantly to the eigenvector normalization. The $1/N$ scaling of the inverse participation ratio shows that the number of such components is of the order of $N$ and therefore the eigenvectors of GUE matrices can be considered as extended over a one-dimensional lattice of size $N$.  This property of the eigenvectors is closely related to the invariance of the distribution function of $H$ under an arbitrary unitary transformation, which in turn follows from the equivalence of the variances $\avg{|H_{ij}|^2}$.

Later on it was realized that RMT can be used in order to describe various phenomena discovered in quantum disordered systems such as  Anderson localization and the metal-insulator transition \cite{Abrahams}. New non-invariant ensembles were introduced, for which the variance of the matrix elements is not a constant, but given by some non-trivial matrix  $\avg{|H_{ij}|^2}=F_{ij}$. Finding statistical properties of the eigenvectors for a generic matrix $F_{ij}$ is too complicated problem to be solved analytically, but some results have been obtained for a certain non-invariant ensembles such as banded random matrices \cite{FM91, FM94}, power law banded random matrices \cite{MFDQS96, ORC11}, almost diagonal random matrices \cite{EM00, KOY11}, ultrametric \cite{FOR09} and Ruijsenaars-Schneider ensembles \cite{BG11} .   

The eigenvectors of such ensembles can be not only extended, but also localized or critical. The eigenvectors of different characters can be distinguished by  scaling of their moments $I_q=\sum_{n=1}^N\avg{|\psi_n|^{2q}}$, which generalize the notion of the inverse participation ratio, with the matrix size $N$:
\be\label{fractal-dim-def}
I_q\propto N^{-d_q(q-1)}.
\en
For GUE matrices Eq.(\ref{gue_distr}) yields $\avg{y^q}=\Gamma (q+1)$, where $\Gamma (z)$ is the gamma function, and hence
\be
I_q^{GUE}=\Gamma (q+1) N^{1-q}.
\en
The above result means that $d_q=1$ and this is the condition, which characterizes extended eigenvectors. For a localized eigenvector, $I_q$ is not sensitive to changing of $N$ and therefore $d_q=0$. If $0<d_q<1$, then one deals with critical states, which are typical at the point of the Anderson metal-insulator transition \cite{EM08}. Their fractal nature can be quantified by the fractal dimensions $d_q$, which can be different for different values of $q$ in the case of multifractal eigenvectors. 

In this work we study the random matrix ensembles, in which the variances of $H_{ij}$ are given by a rank one matrix $\avg{|H_{ij}|^2}=w_i^2w_j^2/N$, where all $w_i>0$. Any matrix  $H$ from such an ensemble can be represented as $H=W\tilde{H}W$, where $\tilde{H}$ is a GUE matrix and $W$ is a diagonal matrix $W_{ij}=w_i\delta_{ij}$, so that this ensemble can be considered as a deformed GUE. The presence of the matrix $W$ breaks the unitary invariance of GUE and provides a preferred basis, in which $W$ is diagonal, making it possible for  localization of eigenvectors to occur.

The deformed GUE  defined in this way is closely related to the generalized eigenvalue problem for a GUE matrix $\tilde{H}$ and a positive definite Hermitian matrix $A$:
\be\label{gen_eigenvalue}
\tilde{H}g=E Ag,
\en
where $E$ is an eigenvalue and $g$ is the corresponding eigenvector \cite{DFF81}. Using the fact that $A$ is positive definite, we can introduce a vector $f=A^{1/2}g$ and rewrite Eq.(\ref{gen_eigenvalue}) as
\be\label{deform_eigenvalue}
A^{-1/2}\tilde{H}A^{-1/2}f=E f.
\en
The unitary invariance of $\tilde{H}$ allows us to choose the basis, in which $A$ is diagonal, and hence the generalized eigenvalue problem (\ref{gen_eigenvalue})  is equivalent to the standard eigenvalue problem for the matrix $W\tilde{H}W$ with $W=A^{-1/2}$. The generalized eigenvalue problem for random matrices appears naturally in various applications such as random reactance networks \cite{F99}, statistical signal processing \cite{NE08}, vibration analysis \cite{S03} and others.  

The mean eigenvalue density for the eigenvalue problem (\ref{gen_eigenvalue}) or (\ref{deform_eigenvalue}), in the case when $H$ and $W$ are both random, was studied a long time ago \cite{P72,G95}. More recently, correlation properties of the eigenvalues were investigated in Ref.\cite{F99}. Remarkably, it was found that the presence of the matrix $W$ does not change the two-point spectral correlation function, which remains the same as for GUE matrices and is given by the well known  Wigner-Dyson expression. At the same time, we demonstrate that depending on the matrix $W$ the eigenvectors of Eq.(\ref{deform_eigenvalue}) can be extended, localized or critical. Thus the random matrices considered in this work is an unusual example of an ensemble, for which localization of the eigenvectors can coexist with the Wigner-Dyson statistics of the eigenvalues, which is typical for extended states.   

The paper is organized as follows. In Section \ref{sec_general_formula} we derive a general result for the moments of the eigenvectors and their distribution function for a generic matrix $W$ using the supersymmetry technique \cite{Efetov}. Various particular choices of the matrix $W$ leading to extended, localized or critical eigenvectors are considered in Section \ref{sec_particular_results}. We calculate analytically the scaling of the moments and confirm our results by numerical simulations. Finally we conclude the paper with a summary of the main results and a discussion of open problems.

\section {General result for the moments of the eigenvectors and the distribution function} \label{sec_general_formula} 
In this section, we derive a general expression for the moments of the eigenvectors of $H= W\tilde{H}W$ for an arbitrary diagonal $W>0$,
using the supersymmetry approach.

The local moments of the $n$th eigenvector component $\psi_n$ at a given energy $E$ are defined as
\be
I_q(n)=\frac{1}{\rho(E)}\sum_{\alpha}\avg{|\psi_n^{\alpha}|^{2q}\delta(E-E_{\alpha})},
\en
where $E_{\alpha}$ is an eigenvalue of $H$ corresponding to a normalized eigenvector $\psi^{\alpha}$ and $\rho(E)=\frac{1}{N}\sum_{\alpha}\avg{\delta (E-E_{\alpha})}$ is the average density of states. 
The moments of the eigenvectors can be expressed through the diagonal matrix elements of  the Green's functions as
\begin{equation}\label{mom-green}
I_q(n) = \frac{\mathrm{i}^{l-m}}{2\pi \rho(E)N} \frac{(l-1)!(m-1)!}{(l+m-2)!}\lim_{\epsilon \to 0}(2\epsilon)^{l+m-1}\avg{(G^R_{nn})^l(G^A_{nn})^{m}},
\end{equation}
where $l$ and $m$ are positive integers such that $l+m=q$ and the retarded $G^{R}$ and advanced $G^{A}$ Green's functions are defined as
\begin{equation}
G^{R/A}(E) = (E\pm \mathrm{i}\epsilon -H)^{-1}.
\end{equation}
The averaged products of the matrix elements of the Green's functions can be calculated effectively using the supersymmetry approach.
The main steps of the method include representing the Green's functions by the Gaussian integrals over supervectors, averaging over the random matrix $\tilde{H}$, decoupling the resulting integral by the Hubbard-Stratonovich transformation, which allows us to integrate out the 
initially  introduced supervectors. All these steps are standard and exactly the same as in the case of GUE, and their detailed  description can be found in the existing literature \cite{Efetov, Haake, F99}. Performing them, we arrive at the following expression:
\begin{align}\label{Q-int} 
\avg{(G^R_{nn})^l(G^A_{nn})^{m}} =&  \int \mathrm{d}\hat{Q} \sum_j C^l_j C^m_j
\left(g_{rr}^{BB}\right)^{l-j}\left(g_{aa}^{BB}\right)^{m-j}\left(g_{ar}^{BB}\right)^j\left(g_{ra}^{BB} \right)^j\\ \notag 
&\times\mathrm{exp} \left[ -\frac{N}{2} \mathrm{Str}\hat{Q}^2 - \sum_{i=1}^N \mathrm{Str\ln}L\{\mathrm{i}E\bm{\hat{1}}-\mathrm{i}\textit{v}_i\hat{Q}-\epsilon\hat{\Lambda}\}\right],  v_i = w_i^2.
\end{align}
where  $C^p_q=p!/(q!(p-q)!)$, the supermatrix $\hat{Q}$ is the integration variable of  the Hubbard-Stratonovich transformation,
$g^{BB}_{\alpha \beta}$ are the matrix elements of the Bose-Bose block of the supermatrix 
$(E\hat{1}-v_n\hat{Q}_n+\mathrm{i}\epsilon\hat{\Lambda})^{-1}$ in the retarded-advanced notation, and
\begin{equation}
 L = \mathrm{diag}(1,1,-1,1), \quad \hat{\Lambda} = \mathrm{diag}(1,1,-1,-1).
\end{equation}

One can notice that in contrast to the standard GUE case the action depends on $v_i$, whereby we can recover the GUE case if we set $v_i = 1$. In the limit $N \rightarrow \infty$, the integral over $\hat{Q}$ is dominated by the saddle-points that satisfy the relation \cite{F99}
\begin{equation}
\hat{Q} = \frac{1}{N}\sum_{i=1}^N \frac{v_i}{E\bm{\hat{1}}-v_i\hat{Q}}.
\end{equation}
The saddle-point solutions can be parametrized as
\begin{equation}
\hat{Q}_{s.p} = t\bm{\hat{1}}-\mathrm{i}s\hat{T}^{-1}\hat{\Lambda}\hat{T},\label{saddlepointeqn}
\end{equation}
where $s \neq 0$ and $t$ are real parameters  satisfying the simultaneous equations
\begin{align}\label{ts-system}
t = \frac{1}{N} \sum_i^N \frac{v_i(E-v_it)}{(E-v_it)^2+s^2v_i^2},\ \ \ \text{and}\ \ \ 1 =\frac{1}{N}\sum_i^N \frac{v_i^2}{(E-v_it)^2+s^2v_i^2},
\end{align}
and the matrix $\hat{T}^{-1}\hat{\Lambda}\hat{T}$ belongs to the standard coset space appearing in the GUE case \cite{Efetov, Haake}. 
As it was shown in Ref.\cite{F99}, using the supersymmetric approach for calculation of the density of states one obtains
\begin{equation}\label{dos}
\rho(E) = \frac{s}{\pi N}\sum_i^N \frac{v_i}{(E-v_it)^2+s^2v_i^2}.
\end{equation}
It turns out that this expression along with the system of equations (\ref{ts-system}) can be considered as a particular case of some general result by Pastur and Girko derived a long time ago \cite{P72, G95}.
 
Applying the saddle-point approximation to the integral (\ref{Q-int}), we find 
\begin{align}
\avg{(G^R_{nn})(G^A_{nn})^{q-1}} = &\int \mathrm{d} \mu (T)  (g_{aa}^{BB})^{q-2}\{g_{aa}^{BB}g_{rr}^{BB}+(q-1)g_{ar}^{BB}g_{ra}^{BB}\}  \notag \\& \times
\mathrm{exp}\left\lbrace -\epsilon\pi N \rho(E)\mathrm{Str}[\hat{T}^{-1}\hat{\Lambda}\hat{T}\hat{\Lambda}]\right\rbrace,
\end{align}
where, for simplicity, we set $m=q-1$ and $l=1$.

In order to explicitly evaluate the integral we have above, over the coset space parametrized by $\hat{T}$, we employ  Efetov's parametrization \cite{Efetov, M99}:
\begin{align}\label{g-implicit}
\avg{(G^R_{nn})(G^A_{nn})^{q-1}} = &\int \mathrm{d} \mu (T) (g_{aa}^{BB})^{q-2}\{g_{aa}^{BB}g_{rr}^{BB}+(q-1)g_{ar}^{BB}g_{ra}^{BB}\}   \notag \\& \times  \mathrm{exp}\left[  -2 \epsilon \pi N \rho(E)(\lambda_1 - \lambda_2) \right],
\end{align}
with $\lambda_1\in [1,\infty)$, $\lambda_2\in [-1,1]$ and the expressions for the integration measure $\mathrm{d}\mu (T)$ and the matrix elements $g^{BB}_{\alpha \beta}$ are given explicitly in \ref{app-parametrization}.

As the action depends only on the parameters $\lambda_1$ and $\lambda_2$, all other variables can be easily integrated out:
\begin{align}\label{lambda-int}
&\avg{(G^R_{nn})(G^A_{nn})^{q-1}}=(q-1)\int_1^{\infty} \mathrm{d} \lambda_1 \int_{-1}^1 \mathrm{d} \lambda_2 \dfrac{(E-v_nt+\mathrm{i}sv_n\lambda_1)^{q-2}}{[(E-v_nt)^2+s^2v_n^2]^q} s^2v_n^2  
\nonumber \\
&\times\Bigg\{1+\dfrac{\lambda_1+\lambda_2}{\lambda_1-\lambda_2}+\dfrac{\mathrm{i}sv_n(q-2)(\lambda_1^2-1)}{(E-v_nt+\mathrm{i}sv_n\lambda_1)(\lambda_1-\lambda_2)}\Bigg\} 
\exp\left[ -2\epsilon \pi N \rho(E)(\lambda_1-\lambda_2)\right].
\end{align}
According to Eq.(\ref{mom-green}) the product of the Green's functions $\avg{(G^R_{nn})(G^A_{nn})^{q-1}}$ is singular 
in the limit $\epsilon \rightarrow 0$ and we need to extract the most singular part of it, in order to calculate the moments of 
the eigenvectors. As $\epsilon$ enters only in the exponential function in the integral in the combination $\epsilon (\lambda_1-\lambda_2)$, it is clear that main contribution to the integral in the limit  $\epsilon \to 0$ is given by large values of non-compact variable $\lambda_1$. This observation motivates a formal change of the variable $\lambda_1$ by $z=\epsilon \lambda_1$. Then the limit $\epsilon \to 0$ can be calculated explicitly, as it is shown in \ref{app-epsilon}.

Once the integration has been completed this leads to our final result for the moments of the eigenvectors
\begin{equation}\label{mom-result}
I_q(n) = \frac{1}{(\pi \rho (E) N)^q} \left[\frac{sv_n}{(E-v_nt)^2+s^2v_n^2}\right]^q \Gamma (q+1).
\end{equation}
In contrast to the GUE case, we notice that the moments are not independent of $n$, therefore each component of the eigenvectors is distributed differently. This is a natural consequence of breaking of the unitary invariance of GUE. Although $I_q(n)$ depends explicitly only on the corresponding value of $v_n$ in a simple way, one should remember that there is a implicit and non-trivial dependence on all $v_i$'s through the variables $s$, $t$ and $\rho(E)$.

Having the result for the moments at our disposal, we can restore the full distribution function of the eigenvectors components:
\begin{align}\label{distr-result}
P_n(x) = \pi \rho(E) N\left[\frac{(E-v_nt)^2+s^2v_n^2}{sv_n}\right]\exp{\left[-(\pi \rho(E)N)\frac{(E-v_nt)^2+s^2v_n^2}{sv_n}x\right]},
\end{align}
where $x = |\psi_n|^{2}$. Eq.\eqref{mom-result} and Eq.\eqref{distr-result} represent the main result of our work. 

To corroborate this result we re-derive the GUE case by setting $v_i = 1$.  As all $v_i = 1$ the system \eqref{ts-system} can be easily solved giving $s = \sqrt{1-(E-t)^2}$ and $t = E/2$. Substituting these expressions into Eq.\eqref{dos} we find
\begin{equation}
\rho (E)  = \frac{1}{\pi} \sqrt{1-(E/2)^2}.
\end{equation}
This is exactly Wigner's semi-circle law for the density of states in the GUE case. Substituting the density of states and the expressions for $s$ and $t$  into our result for the moments we see that it reproduces exactly  the GUE result given in the introduction:
\begin{equation}
\sum_n^N I_q(n) = \Gamma(q+1)N^{1-q}.
\end{equation}
We also note as a special case the moments for $E=0$ valid for any $v_i$,
\begin{equation}
I_q(n) = \frac{v_n^{-q}}{(\sum_i \frac{1}{v_i})^q} \Gamma(q+1) \label{zeroenergy}
\end{equation}
with the corresponding distribution function as
\begin{equation}
P_n(x) = v_n\sum_i^N v_i^{-1}\exp{\left[-v_n\sum_j^N v_j^{-1}x\right]}.
\end{equation}
Eq.\eqref{zeroenergy} can be interpreted as follows. The eigenvectors of $W\tilh W$ can be obtained from the GUE eigenvectors of  $\tilh$ simply by multiplying their $n$th component by $v_n^{-1/2}=w_n^{-1}$ and then normalizing them. The existence of such a simple interpretation suggests that there is probably a more direct way of getting this result. It turns out that this is indeed the case, as we show in the next section. 

\section {Alternative derivation of the zero energy result for the moments of the eigenvectors} \label{sec_alternative_derivation}

In this section we present an alternative derivation of the zero energy result (\ref{zeroenergy}), which does not involve the supersymmetry technique. Instead it is based on the well known result for the distribution function of the eigenvector components of GUE matrices and the fact that there is a straightforward relation between the eigenvectors of $H$ and $\tilde{H}$ corresponding to $E=0$.

The zero energy eigenvector $g$ of $H$ satisfies $W\tilh Wg=0$, which is equivalent to $\tilh Wg=0$, as $w_i\neq 0$.  Hence $f\equiv Wg$ is the zero energy eigenvector of $\tilh$. If we assume that $f$ is normalized then $g=W^{-1}f$ is not normalized, and we define its normalized counterpart $u\equiv g/||g||$. The components of $u$ read
\be
u_n=\frac{w_n^{-1}f_n}{\left(\sum_i|w_i^{-2}f_i^2|\right)^{\frac{1}{2}}},
\en
so that any component of $u$ depends on all components of $f$ due to the normalization condition. The averaged local moments of the eigenvector $u$  are given by
\be
I_q(n)\equiv\avg{|u_n|^{2q}}=\avg{\frac{|w_n^{-1}f_n|^{2q}}{\left(\sum_i|w_i^{-2}f_i^2|\right)^{q}}}.
\en
The main difficulty in computing the above average comes from the term in the denominator. In order to overcome this problem we use the following integral representation for the denominator
\be
\left(\sum_i|w_i^{-2}f_i^2|\right)^{-q}=\frac{1}{\Gamma(q)}\int_0^{\infty}d\alpha \:\alpha^{q-1}e^{-\alpha \sum_i x_i |f_i|^2},
\en
where $x_i\equiv w_i^{-2}$ and $\Gamma(z)$ is the  gamma function. Then the expression for $I_q(n)$ can be written as
\be
I_q(n)=\frac{1}{\Gamma(q)}\int_0^{\infty}d\alpha \:\alpha^{q-1}x_n^q\avg{|f_n|^{2q}e^{-\alpha \sum_i x_i |f_i|^2}}.
\en
It is well known that the components of a normalized eigenvector of GUE matrix become statistical independent in the limit $N\to \infty$ and the distribution function of $y\equiv N|f_i|^2$ is given by $P(y)=e^{-y}$. Using this result the averaging over $f_i$ can be easily calculated:
\be\label{mom_int_alpha}
I_q(n)=\frac{\Gamma(q+1)}{\Gamma(q)}\left(\frac{x_n}{N}\right)^q\int_0^{\infty}d\alpha \:
\alpha^{q-1}\left(1+\frac{\alpha x_n}{N}\right)^{-q}\prod_i \left(1+\frac{\alpha x_i}{N}\right)^{-1}.
\en  
The above integral can be calculated asymptotically in the limit $N\to \infty$. Keeping only the leading in $1/N$ terms we can write the integral over $\alpha$ as
\be\label{int_alpha}
\int_0^{\infty}d\alpha \:\alpha^{q-1}e^{-\frac{\alpha}{N}\sum_ix_i}=\left(\frac{1}{N}\sum_ix_i\right)^{-q}\Gamma(q),
\en 
which yields the final result for the local moments
\be\label{final_res}
I_q(n)=\frac{x_n^{q}}{\left(\sum_i x_i\right)^q}\Gamma(q+1).
\en
Recalling that  $x_i=w_i^{-2}=v_i^{-1}$, we can see that this result is indeed equivalent to Eq.(\ref{zeroenergy}) derived in the previous section. We would like to stress that the above derivation works only for $E=0$, where a simple relation between the eigenvectors of the two problems can be established.

This approach allows us also to establish  a simple necessary condition for the validity of the final result. Indeed, deriving Eq.\eqref{int_alpha} from the integral in Eq.\eqref{mom_int_alpha} we assumed that $\sum_i (x_i/N)^2\ll \sum_i x_i/N$. Taking into account that 
$\rho (0)=(1/\pi N) \sum_i x_i$, we conclude that a necessary condition reads
\be\label{condition}
\sum_i x_i^2\ll \rho (0)N^2.
\en 

\section {Extended, localized and critical eigenvectors} \label{sec_particular_results} 
In this section we consider few specific choices for entries of the diagonal matrix $W$, which enable us to see how the deformation of the ensemble changes the nature of the eigenvectors. In particular, we find that depending on choice of $W$ the eigenvectors of  $H=W\tilh W$
can be extended, localized or critical.

The zero energy formula \eqref{zeroenergy} provides the convenient starting point of our analysis due to its simplicity. However it turns out that scaling of the moments in the limit $N \to \infty$ remains qualitatively the same regardless of the energy. Therefore we present here analytical results in the generic case for arbitrary value of $E$. We verify them by direct numerical diagonalization of random matrices in two different cases.  

We focus on the power-law dependence of $v_n$ on $n$, which as we show below is an interesting example comprising a variety of different types of eigenvectors. 
\begin{equation}
v_n = c\left(\frac{1}{n}\right)^p,
\end{equation}
where $c$ is the normalization constant, chosen in such a way that the density of states is independent of $N$ as $N \rightarrow \infty$.
It must be split into three cases, $p>0$, $-1<p<0$, and $p<-1$, as the scaling of the moments is different in each case.

In the case of $p>0$, $v_n$ must be normalized as shown below
\begin{equation}
v_n = \left(\frac{N}{n}\right)^p, \ p>0.
\end{equation}
Substituting $v_n$ into the expression for the moments we obtain
\begin{equation}\label{power-law-gen}
I_q\equiv\sum_n^N I_q(n) = \frac{c^{-q}}{N^q} \sum_n^N \left[\frac{s(N/n)^p}{(E-(N/n)^pt)^2+s^2(N/n)^{2p}} \right]^q \Gamma (q+1).
\end{equation}
Estimating the divergent sum in the above equation, we find that the scaling of the moments  in this case is independent of $p$ and is given by
\begin{equation}
I_q \propto N^{1-q} \label{27}.
\end{equation}
The scaling is the same as in the GUE case and it corresponds  to the extended states. Thus this particular deformation of the ensemble has no qualitative effect on the nature of the eigenvectors.

For the next case, that we consider,  $-1<p<0$,  the normalization constant is the same as in the previous one:
\begin{equation}
v_n = \left(\frac{N}{n}\right)^p,\ -1<p<0 .
\end{equation}
The expression for the moments is formally the same as in Eq.\eqref{power-law-gen}, however for $-1<p<0$ the sum can converge or diverge depending on value of $q$. As a result we obtain
\begin{align}\label{mom_second}
 I_q\propto  \begin{cases}
N^{-q(p+1)}, & q>-\frac{1}{p}, \\
N^{1-q}, & q<-\frac{1}{p},\\
 \ln(N) N^{1-q}, & q=-\frac{1}{p}.
\end{cases}
\end{align}
The scaling with the non-trivial power of $N$ corresponds to the critical states, whose fractal dimensions can be determined by comparison of the above result with Eq.\eqref{fractal-dim-def}:
\be
d_q=
\begin{cases}
\frac{q(p+1)}{q-1}, & q>-\frac{1}{p}, \\
1, & q<-\frac{1}{p} .
\end{cases}
\en
This result indicates that the eigenvectors belong to the sort of ``frozen" phase \cite{EM08} and combine properties of extended and critical states.

Finally we consider the case  $p<-1$, in which the normalization constant must be altered and  $v_n$ is given by 
\begin{equation}
v_n = \frac{1}{Nn^p},\ p < {-1}.
\end{equation}
As the convergence of the sum in Eq.\eqref{power-law-gen} is again determined by the value of $pq$, three different sub-cases $pq<-1$, $pq=-1$, and $pq>-1$ must be considered separately \footnote{The necessary condition \eqref{condition} is not directly applied to this case. However, one can see that the integral over $\alpha$ in Eq.\eqref{mom_int_alpha} becomes $N$ independent and it tends to a constant, as $n\to\infty$. As a result $I_q(n)\propto (x_n/N)^q$, which yields the same scaling with $N$ as Eq.\eqref{final_res} and Eq.\eqref{mom-result}.}. Estimating the sum in each of these cases we find
\begin{align}
I_q\propto \begin{cases}
\mbox{const}, & q>-\frac{1}{p}, \\
N^{pq+1}, & q<-\frac{1}{p}, \\
\ln(N), & q=-\frac{1}{p},
\end{cases}
\end{align}
and the corresponding fractal dimensions read
\be
d_q=
\begin{cases}
0, & q>-\frac{1}{p}, \\
\frac{pq+1}{1-q}, & q<-\frac{1}{p} .
\end{cases}
\en
This time the eigenvectors belong to a different ``frozen" phase, in which eigenstates share properties of localized and critical states.

\begin{figure}[t]\label{fig_Iq}
\centering
\includegraphics[width=0.8\textwidth]{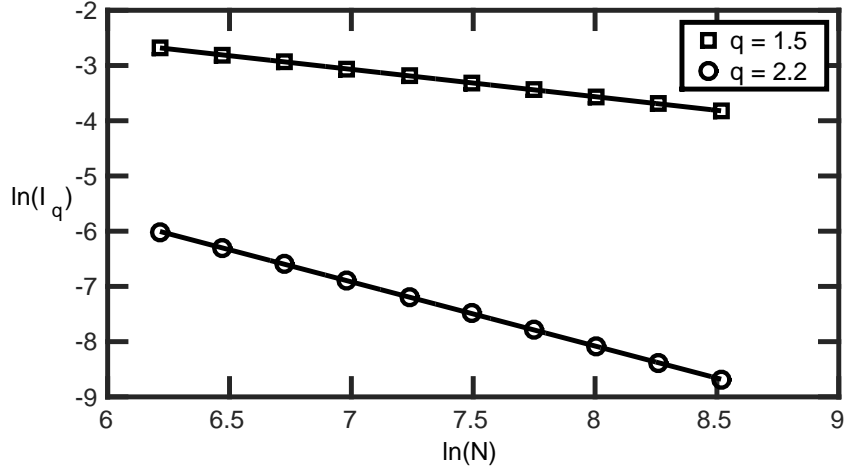}
\caption{Numerical results (symbols)  for $\ln (I_q)$ as a function of $\ln (N)$ for q = 1.5 and q = 2.2. The solid lines represent the best fit to the numerical data.}
\end{figure}

Numerical simulations using direct diagonalization  were performed to test the validity of the analytical results for $v_n = (N/n)^{-1/2}$, with $N$ ranging from 500 to 5000. The moments were calculated over 5000 realizations for the eigenvectors, for which the corresponding eigenvalues were close to $E = 0$. The numerical results for  $q = 1.5$, and $q = 2.2$ are presented in Fig.~\ref{fig_Iq} along with the best fit solid lines. The numerical values of the gradients of the lines are $-0.49$ and $-1.16$ for $q = 1.5$ and $q = 2.2$ respectively, which is in nice agreement with the analytical results $-0.5$ and $-1.1$ following from Eq.\eqref{mom_second}.

Finally we discuss briefly the possibility of having completely localized states for the deformed Hamiltonian $H$. The zero energy result \eqref{zeroenergy} suggests that such states appear in our model provided that the sequence $\{v_n\}$ is itself localized in space.  For example, taking $v_n= Nx^{-n}$ with $x>1$, we obtain that all the moments $I_q\to \mbox{const}$ for all $q>0$, as $N\to \infty$, implying that the eigenvectors are localized in this case.

\section {Conclusions} \label{sec_conclusions} 

In this paper we introduced the non-invariant matrix ensembles, in which the variances of the random matrix elements are position dependent and given by  $\avg{|H_{ij}|^2}=w_i^2w_j^2/N$. We studied  the eigenvector statistics of such matrices  in the limit $N \to \infty$ and found that the local statistics is determined by $v_i=w_i^2$. For the eigenvectors corresponding to the zero eigenvalue $E=0$ the expression for the moments of the eigenvectors takes a particular simple form \eqref{zeroenergy}. The general result \eqref{mom-result}, which is valid for arbitrary $E$, was derived using the supersymmetry technique. It turns out that two parameters $s$ and $t$, which enter in 
Eq.\eqref{mom-result}, are determined by the same systems of coupled equations as appeared first in the work Pastur and Girko, who studied the density of eigenvalues  for deformed random matrix ensembles. The full information of the eigenvectors statistics is given by the distribution function of their components, which we found in Eq.\eqref{distr-result}.

Our general result can be applied to any particular choice of $v_n$. We considered in detail the power law dependence $v_n\propto 1/n^p$. In this case we showed that, by varying $p$, eigenvectors are changing from extended, for $p>0$, to critical quasi-extended, for $-1<p<0$, and further to critical quasi-localized, for $p<-1$. Other choices of $v_n$ may lead to completely localized states, such as, for example, the exponential dependence $v_n\propto x^{-n}$ with $x>1$.

It would be interesting to consider a similar problem for the matrix $\tilh$ from other symmetry classes. Our calculations and especially the alternative derivation of the zero energy result suggest that the formula for the moments will have similar structure for other symmetry classes: it will contain the universal factor, describing the dependence on $v_n$, such as the factor $v_n^{-q}(\sum_i v_i^{-1})^{-q}$ in Eq.\eqref{zeroenergy}, and a symmetry dependent factor, which is the same as for the non-deformed  ensemble, such as $\Gamma(q+1)$ for GUE.

We thank Yan Fyodorov for useful comments. KT acknowledges support from the Engineering and Physical Sciences Research Council [grant number EP/M5065881/1]. 

\appendix
\section{Pre-exponential factors in Efetov's parametrization}\label{app-parametrization}

The pre-exponential factors entering into Eq.\eqref{g-implicit} are given in  Efetov's parametrization by the following expressions: 
\begin{eqnarray}
g_{aa}^{BB} &=& \dfrac{E-v_nt+\mathrm{i}sv_n\lambda_1 + \mathrm{i}sv_n (\lambda_1 - \lambda_2) \alpha \alpha^*}{(E-v_nt)^2+s^2v_n^2} ,\\
g_{ar}^{BB} &=& -\dfrac{\mu_1sv_n\left(1+\dfrac{\alpha \alpha^*}{2}\right)\left(1-\dfrac{\beta \beta^*}{2}\right)+\mu_2^*sv_n\alpha^* \beta}{(E-v_nt)^2+s^2v_n^2}, \\
g_{ra}^{BB} &=& -\dfrac{\mu_1^*sv_n\left(1-\dfrac{\beta \beta^*}{2}\right)\left(1+\dfrac{\alpha \alpha^*}{2}\right) + \mu_2sv_n\beta^*\alpha}{(E-v_nt)^2+s^2v_n^2}, \\
g_{rr}^{BB} &=& \dfrac{E-v_nt - \mathrm{i}sv_n \lambda_1+\mathrm{i}sv_n(\lambda_1-\lambda_2)\beta \beta^*}{(E-v_nt)^2+s^2v_n^2}.
\end{eqnarray}

The expression for the integration measure reads
\begin{equation}
\mathrm{d}\mu(T) = -\dfrac{\mathrm{d}\lambda_1 \mathrm{d} \lambda_2}{(\lambda_1-\lambda_2)^2} \mathrm{d}\phi_1 \mathrm{d}\phi_2 \mathrm{d} \alpha \mathrm{d} \alpha^*\mathrm{d} \beta \mathrm{d} \beta^*.
\end{equation} 
where $\lambda_1 \in [1,\infty),\ \lambda_2 \in [-1,1],\ \phi_2,\phi_2 \in [0,2\pi]$, and $\alpha, \alpha^*, \beta, \beta^*$ are Grassmann variables, for which the following convention is used  
\begin{equation}
\int \mathrm{d}\alpha\ \alpha =\int \mathrm{d}\alpha^*\ \alpha^* =\int \mathrm{d}\beta\ \beta =\int \mathrm{d}\beta^*\ \beta^* =\frac{1}{\sqrt{2 \pi}}.
\end{equation}

\section{Evaluation of the integral over $\lambda_1$ and $\lambda_2$}\label{app-epsilon}
In order to evaluate the integral in Eq.\eqref{lambda-int} in the limit $\epsilon\to 0$ we make use of the substitution $z = \epsilon \lambda_1$:
\begin{align}
&\avg{(G_{nn}^R) (G_{nn}^A)^{q-1}}=(q-1)\int_1^{\infty} \mathrm{d}\left(\frac{z}{\epsilon}\right) \int_{-1}^1 \mathrm{d} \lambda_2 \dfrac{(E-v_nt+\mathrm{i}sv_nz/\epsilon)^{q-2}}{[(E-v_nt)^2+s^2v_n^2]^q}s^2v_n^2 \notag \\ &\times \Bigg\{1+ \dfrac{(z/\epsilon+\lambda_2)}{z/\epsilon-\lambda_2}\notag +\dfrac{\mathrm{i}sv_n(q-2)(z^2/\epsilon^2-1)}{(E-v_nt+\mathrm{i}sv_nz/\epsilon)(z/\epsilon-\lambda_2)}\Bigg\} \mathrm{exp}\left\lbrace -2\epsilon \pi N \rho(E)(z/\epsilon-\lambda_2)\right\rbrace.
\end{align}
Extracting the most singular part of the above expression we find that the $\lambda_2$ dependence in the integrand vanishes and as a result we obtain
\be
\avg{(G_{nn}^R) (G_{nn}^A)^{q-1}}=\frac{2q(q-1)\mathrm{i}^{q-2}\epsilon^{1-q}(sv_n)^q}{[(E-v_nt)^2+s^2v_n^2]^q}
\int_0^{\infty} d z z^{q-2} \mathrm{exp}\left\lbrace - 2 \pi \rho(E) N z \right\rbrace +O\left(\epsilon^{2-q}\right).\nonumber
\en
Calculating the integral over $z$ and substituting the result into Eq.\eqref{mom-green} we arrive at Eq.\eqref{mom-result}.


\end{document}